\documentclass[12pt]{article}

\newsavebox{\ns}
\newsavebox{\dbrane}
\newsavebox{\dbshort}

\usepackage{epsfig}
\usepackage{amssymb}

\def\appendix{{\newpage\section*{Appendix}}\let\appendix\section%
        {\setcounter{section}{0}
        \gdef\thesection{\Alph{section}}}\section}

\def\be{\begin{eqnarray}}
\def\ee{\end{eqnarray}}

\newcommand{\nn}{\nonumber}
\newcommand\para{\paragraph{}}
\newcommand{\ft}[2]{{\textstyle\frac{#1}{#2}}}
\newcommand{\eqn}[1]{(\ref{#1})}

\def\Dslash{\,\,{\raise.15ex\hbox{/}\mkern-12mu D}}
\def\Dbarslash{\,\,{\raise.15ex\hbox{/}\mkern-12mu {\bar D}}}
\def\delslash{\,\,{\raise.15ex\hbox{/}\mkern-9mu \partial}}
\def\delbarslash{\,\,{\raise.15ex\hbox{/}\mkern-9mu {\bar\partial}}}
\def\pslash{\,\,{\raise.15ex\hbox{/}\mkern-9mu p}}
\def\calDslash{\,\,{\raise.15ex\hbox{/}\mkern-12mu {\cal D}}}

\input amssym.def
\input amssym.tex

\def\bbox{{\,\lower0.9pt\vbox{\hrule \hbox{\vrule height 0.2 cm
\hskip 0.2 cm \vrule height 0.2 cm}\hrule}\,}}

\def\Dslash{\,\,{\raise.15ex\hbox{/}\mkern-12mu D}}
\def\Dbarslash{\,\,{\raise.15ex\hbox{/}\mkern-12mu {\bar D}}}
\def\delslash{\,\,{\raise.15ex\hbox{/}\mkern-9mu \partial}}
\def\delbarslash{\,\,{\raise.15ex\hbox{/}\mkern-9mu {\bar\partial}}}
\def\pslash{\,\,{\raise.15ex\hbox{/}\mkern-9mu p}}
\def\calDslash{\,\,{\raise.15ex\hbox{/}\mkern-12mu {\cal D}}}

\begin{document}
\pagestyle{plain}
\setcounter{page}{1}
\newcounter{bean}
\baselineskip16pt

\begin{titlepage}

\begin{center}
\today
\hfill hep-th/0310221\\
\hfill MIT-CTP-3426 \\
\hfill SLAC-PUB-10211\\
\hfill SU-ITP-3/29 \\
\hfill NSF-KITP-03-87 \\

\vskip 1.5 cm
{\Large \bf Scalar Speed Limits and Cosmology: \\ Acceleration from D-cceleration}
\vskip 1 cm
{Eva Silverstein${}^1$ and David Tong$^{2}$}\\
\vskip 1cm {\sl ${}^1$SLAC and Department of Physics, Stanford
University, \\Stanford, CA 94309/94305, U.S.A. \\ {\tt
evas@slac.stanford.edu} } \vskip .3cm {\sl ${}^2$ Center for
Theoretical Physics,
Massachusetts Institute of Technology, \\ Cambridge, MA 02139, U.S.A. \\
{\tt dtong@mit.edu}}

\end{center}

\vskip 0.5 cm
\begin{abstract}

Causality on the gravity side of the AdS/CFT correspondence
restricts motion on the moduli space of the ${\cal N}=4$ super
Yang Mills theory by imposing a speed limit on how fast the scalar
field may roll. This effect can be traced to higher derivative
operators arising from integrating out light degrees of freedom
near the origin. In the strong coupling limit of the theory, the
dynamics is well approximated by the Dirac-Born-Infeld Lagrangian
for a probe D3-brane moving toward the horizon of the AdS Poincare
patch, combined with an estimate of the (ultimately suppressed)
rate of particle and string production in the system. We analyze
the motion of a rolling scalar field explicitly in the strong
coupling regime of the field theory, and extend the analysis to
cosmological systems obtained by coupling this type of field
theory to four dimensional gravity.  This leads to a mechanism for
slow roll inflation for a massive scalar at subPlanckian VEV
without need for a flat potential (realizing a version of
k-inflation in a microphysical framework). It also leads to a
variety of novel FRW cosmologies, some of which are related to
those obtained with tachyon matter.

\end{abstract}

\end{titlepage}

\section{Introduction}



It is now almost 100 years since Einstein introduced the concept
of a universal speed limit for all physical systems propagating in
spacetime. However, in general, motion on the configuration space
of a physical system is not constrained to obey a speed limit. For
example, the vacuum expectation values (VEVs) of light scalar
fields naturally move in an internal moduli space where motion is
uninhibited; within classical, relativistic field theory there is
no restriction on the rate of change of the VEV. The purpose of
this paper is to show that this situation does not necessarily
continue to hold in the quantum theory. We exhibit situations
where the strong coupling dynamics do impose a speed limit on an
internal moduli space, and examine applications to cosmology,
including a mechanism for slow roll inflation.

\para
We focus on the ${\cal N}=4$ supersymmetric Yang Mills (SYM)
theory. Although the (supersymmetry-protected) metric on the
moduli space is flat, the quantum induced speed limit ensures that
a rolling scalar field slows down as it approaches the origin.
This fact can be seen immediately from the gravity side of the
AdS/CFT correspondence, where the process corresponds to a
D3-brane domain wall in $AdS_5$ moving toward the horizon. The
familiar causal speed-limit in the bulk translates into a speed
limit on moduli space which becomes more pronounced as the brane
approaches the origin. Indeed, from the perspective of a boundary
observer, the probe brane takes an infinite time to cross the
horizon (though as we will discuss, the probe approximation breaks
down due to back reaction at very late times). This speed limit on
the moduli space, arising from causality in the bulk, was first
stressed by Kabat and Lifschytz \cite{kl}.

\para
On the field theory side of the correspondence, this result
reflects the breakdown of the moduli space $\sigma$-model
approximation as a scalar field approaches a locus with new light
degrees of freedom where higher derivative terms become important.
In AdS/CFT dual pairs, we can use the gravity side of the
correspondence to determine the net effect of these
higher-derivative corrections where they are summed into a
Dirac-Born-Infeld (DBI) action. The resulting dynamics is
dramatically different from naive expectation based on the
supersymmetric moduli space metric.

\para
In this paper, we will study the dynamics explicitly, both in
quantum field theory in its own right and in the cosmological
context arising from quantum field theory coupled to four
dimensional gravity. We use the dual picture of $D3$-branes and
anti-$D3$-branes moving in an $AdS$-like throat and we exhibit
late time solutions describing the physics as the scalar field
approaches the origin. Among our results, we find that the slowing
down of the scalar field (relative to the behavior predicted by
the two derivative action) can lead to new regimes exhibiting
inflationary behavior.


\para
The idea of obtaining inflation from $\overline{D3}$ and $D3$
branes in a warped throat was studied recently in \cite{dflation}.
In particular, the authors observe that one can gain extra control
from the warping, but some obstacles to obtaining explicit
inflationary models in string compactifications were identified.
Our results here concern the effects of the crucial higher
derivative terms in the effective action and allow us to probe a
different regime from that studied in \cite{dflation} which may
help address the challenges of \cite{dflation}. In particular the
slow motion of the scalar field, enforced by higher derivative
terms, leads to a new mechanism for slow roll inflation.

\para
A scalar field mass is required to obtain our simplest
inflationary solution.  The required mass scale is not finely
tuned in ordinary four dimensional effective field theory terms,
which provides an interesting distinction from the usual
inflationary models.  In the case of the ${\cal N}=4$ super Yang
Mills theory coupled to gravity, we propose some effective field
theory couplings to other sectors generating such a mass while
leaving intact the form of the crucial kinetic corrections to the
moduli space approximation of the field theory. However, we should
note that the couplings of Kaluza Klein modes in the corresponding
brane throat to other sectors may be important and may require
tuned coefficients to avoid destabilization of the throat; this is
a little understood aspect of current compactification technology.
Because of the plethora of independent ingredients (and types of
domain wall branes) available in string compactifications, we
expect the combination of approximate AdS-metric induced kinetic
terms and scalar masses generated from other couplings is likely
to be available in some set of examples.  It remains the main
weakness of our results however that we will not exhibit an
explicit example here.  The other field theoretic and cosmological
phases we will exhibit are not tied to this subtlety.

\para
Our results have further relations to previous work. The use of
higher derivative terms to change scalar field dynamics in ways
interesting for cosmology has been dubbed ``k-inflation"
\cite{kflation} or ``k-essence" \cite{kessence} .  A particularly
well-studied example occurs in the effective field theory
describing the decay of branes and antibranes leading to ``tachyon
matter" \cite{tachyon,Maloney:2003ck,llm} and the associated
cosmology \cite{tacycosm,others}. While our system shares many features
with the rolling tachyon story, including the existence of a
pressureless dust equation of state, it also differs in several ways
important for cosmology. In particular, we shall argue that
particle and string production is suppressed in our system and the
dynamics is governed to good approximation by our equations of
motion.  For a few examples of previous works studying the
application of rolling moduli to string cosmology see
\cite{Horne:1994mi,Banks:1995dp,Banks:1998vs} and the review
\cite{quevedo}. The crucial physics of our model is ultimately
extracted from the dynamics of D-branes; there are of course many
interesting investigations in this area, for example
\cite{Bachas:1995kx,Douglas:1996yp,Burgess:2003qv}.

\para
The manner in which higher derivative terms can drastically affect
the dynamics of a system raises many questions concerning
potential applications to real models of cosmology as well as more
theoretical issues.  Among the latter set is the question of
whether or not motion toward other finite distance singularities
in field theory and string theory moduli spaces (such as the
conifold singularity) also exhibits similar slowing down effects
for some regime of the couplings. Noticeably, recent studies of
the flop transition, which explore the effects of the
states which become light
at the singularity, suggest that the rolling scalar field does indeed
become stuck in the region of the singularity \cite{lukas,jarv}.

\para
The paper is organized as follows. We start in Section 2 by
describing our basic set-up, stressing the appearance of a speed
limit on the moduli space. In Section 3 we study the consequences
of this speed limit for the dynamics in the global conformal field
theory. We then couple our system to gravity in Section 4. We take
particular care to describe the possible deformations of the
system arising from other sectors in a string compactification,
including the generation of a potential on the moduli space and
the effects this has on the AdS geometry seen by the probe brane.
In Section 5 we study cosmologies arising from our low-energy
effective actions. In Section 6, we check that our solutions are
not destabilised by perturbations or particle production.

\section{The Basic Setup}

Typically studies of scalar field dynamics consider an effective
field theory Lagrangian containing a kinetic term (up to two
derivatives) together with a potential energy on the space of scalar fields:
\be S_{two~deriv}=\int d^4x \sqrt{g}
\left(G_{ij}(\phi)g^{\mu\nu}\partial_\mu\phi^i\partial_\nu\phi^j-
V(\phi)\right)
\label{usualact}\ee
In supersymmetric situations, the moduli space metric
$G_{ij}$ and the form of the potential energy $V$ are highly
constrained, and much of the recent work on such quantum field
theories has focused on determining these quantities and the BPS
spectrum of states exactly.

\para
It is well known however that the moduli space approximation (in
which the physics is governed by the action \eqn{usualact} along
flat directions of the potential $V(\phi)$) can break down due to the
presence of new light degrees of freedom arising on special loci
of the space of scalar field VEVs.

\para
In the $U(N)$ ${\cal N}=4$ SYM theory, one has scalar fields in
the adjoint representation of the gauge group which can be
represented by $N\times N$ matrices $\Phi$.  The moduli space of
the field theory is parameterized by diagonal (commuting)
matrices, with eigenvalues $\phi_1,\dots, \phi_N$ whose moduli
space metric, including quantum corrections, is flat.  Away from
the origin $\phi_i=0$ of this moduli space, the low energy gauge
symmetry is generically $U(1)^N$ and off diagonal matrix elements
of the scalar fields, fermions, and gauge bosons obtain masses. We
shall refer to these modes collectively as ``W bosons". In
particular, we will consider a configuration in which
$\langle\phi_1\rangle\equiv \langle\phi\rangle \ne 0$ but all the
other $\phi_i$, $i\neq 1$ have vanishing VEVs.  This means that
the theory has a low energy unbroken gauge symmetry $U(N-1)\times
U(1)$ with massive W bosons in the $({\bf N-1}, {\bf +})\oplus
({\bf\overline{N-1}},{\bf -})$ representation.  The W bosons are
BPS protected states with mass \be m_W=\phi \label{MW}\ee We will
work expanding about the large $N$ limit of the theory, in which
the natural parameters in the field theory are the rank $N$ and
the 't Hooft coupling $\lambda = g_{YM}^2 N$ in terms of the Yang
Mills coupling $g_{YM}$.

\subsection{The System at Weak Coupling}

At weak coupling in the field theory, the effective action for
$\phi$ gets contributions from virtual W bosons. For large
$\langle\phi\rangle$, these contributions scale like powers
of ${\lambda\dot\phi^2/\phi^4}$ in the planar limit (see
\cite{tsy} for a comprehensive discussion of these higher
derivative corrections). If we send the scalar field rolling
toward the origin from a finite point on the Coulomb branch, the
classical action \eqn{usualact} would predict a constant velocity
$\dot\phi=v_0$, but the form of the corrections just noted shows
that the action \eqn{usualact} becomes unreliable at the distance
\be \phi^2=\sqrt{\lambda}\,v_0 \label{first}\ee The question then
arises about how to take into account these corrections. However,
in the weak coupling regime, the point is moot as we have another
issue to confont before we get this close to the origin. Since the
W bosons become light, they may be produced on shell during the
evolution. The time dependent mass leads to particle production
controlled by the parameter $\dot{m}_W/m_W^2$, which therefore
becomes important at the distance, \be \phi^2=v_0
\label{second}\ee For $\lambda\ll 1$, we first reach the
production point \eqn{second}.

\para
Naively extrapolating the above perturbative results to
the strong coupling regime, we reach the point \eqn{first} first and
therefore expect that the dynamics will be governed by the
effective action for $\phi$ taking into account the
$\lambda\dot\phi^2/\phi^4$ corrections arising from virtual W
bosons. In the following section we shall confirm this expectation
and show that the scale \eqn{first} is where the speed limit on
the moduli space first becomes apparent.

\subsection{The System at Strong Coupling and the Speed Limit}

At strong coupling $\lambda\gg 1$, the effective description of the
theory is via gravity and string theory using the AdS/CFT
correspondence \cite{bigreview}.  A point on the Coulomb branch is described
on the gravity side in the Poincare patch via a D3-brane at fixed
radial position $r$ in the metric
\be ds^2=\frac{r^2}{R^2}\,(-dt^2+dx^2)+\frac{R^2}{r^2}\,dr^2
\label{AdSmetric}\ee
and at a point on the $S^5$.  The field theory coordinate $\phi$
on the Coulomb branch translates into $r/\alpha'$ in the
gravity variables.  We further have the relations
$R=(g_sN)^{1/4}\sqrt{\alpha'}$ and $g^2_{YM}=4\pi g_s$.
\para
As in other applications of AdS/CFT \cite{bigreview} and
Randall-Sundrum \cite{RS}, the warp factor in \eqn{AdSmetric}\
plays an important role in understanding the energy spectrum of
the system. The string mass scale at position $r$ is related to
that at position $r'$ by the ratio of warp factors $r/r'$.  The
open string oscillator modes on the brane at position
$r=\langle\phi\rangle\alpha'$ have masses of order
\be m_s(\langle\phi\rangle)={\phi\over\lambda^{1/4}} \label{MS}\ee
As in the Randall Sundrum scenario, the effective cutoff for modes
on the brane is at this warped string energy scale \eqn{MS} rather
than the UV string scale $1/\sqrt{\alpha'}$.
\para
The W bosons are strings stretched from the brane to the horizon
at $r=0$. Although this is an infinite proper distance for the
string to stretch, the warp factor $r/R$ reduces the effective
tension of the string enough to produce a finite total mass
$m_W=\phi$
in accord with the BPS formula \eqn{MW}.

\para
We would now like to highlight the simple feature of our system
which lies behind most of our detailed results to follow. It is
immediately clear from \eqn{AdSmetric} that, for a boundary
observer using coordinates $(t,x)$, a probe brane takes infinite
time to reach the origin of the Coulomb branch at $r=0$.  (In our
solutions it will turn out that the proper time for the probe to
fall to the origin will be finite.)
The radial velocity of the $D3$-brane in the AdS space is limited
by the speed of light which, translated in field theory variables,
becomes \be \dot\phi\le\dot\phi_c=\frac{\phi^2}{\sqrt{\lambda}}
\label{limit}\ee This restriction was noted previously by Kabat
and Lifschytz \cite{kl}, who discussed some interesting aspects of
the phenomenon. Note that the probe brane takes infinite time
despite the fact that the moduli space metric in \eqn{usualact}
for the ${\cal N}=4$ SYM theory is uncorrected quantum
mechanically; the distance to the origin in the field theory
moduli space metric is finite for any value of the coupling. As
discussed above, the fact that it takes an infinite time to reach
the origin arises from crucial corrections to the moduli space
approximation.  These corrections apply to {\it any} physical
process in which the scalar rolls toward the origin of the Coulomb
branch.

\para
Note that the $\lambda\dot\phi^2/\phi^4$ corrections, while they
are higher derivative corrections, are not suppressed by powers of
the Planck mass; as we discussed above they are suppressed only by
the W boson mass $\phi$.  Let us compare this situation to a brane
in flat space: such an object has a Lagrangian proportional to
$\sqrt{1-v^2}$ where $v$ is the proper velocity of the brane.
Written in terms of canonically scaled brane fields $\phi$, this
becomes (for a D3-brane) $\sqrt{1-\dot\phi^2\alpha'^2}$ which
expands to a series of higher derivative terms which are
suppressed by powers of the string mass $m_s$.  In the global
limit $m_s\to\infty$, these corrections die and the motion on
$\phi$ space is unconstrained.  It is the warp factor in
\eqn{AdSmetric}\ that produces higher derivative effects that are
crucial at the field theory level, as we will see in detail in our
analysis.

\section{The Global CFT}

The corrections to the moduli space approximation, and the
resulting dynamics, can be understood rather explicitly in a
controlled analysis on the gravity side in which the rolling
scalar VEV is modelled by a moving D3-brane probe in
$AdS_5$. In this section we examine the resulting dynamics
of the scalar field.

\subsection{Effective Action and Approximation Scheme}

\para
On the gravity side, the effective Lagrangian appropriate for a
probe D-brane at arbitrary velocity (less than or equal to the
speed of light) but low proper acceleration is the
Dirac-Born-Infeld Lagrangian. This Lagrangian describes the
effects of virtual open strings at the planar level; it includes
the effects of the background geometry and field strengths but
does not include production of on shell closed strings or W bosons
or loops of closed strings. We will analyze the motion of the
3-brane starting from this Lagrangian, and check for self
consistency of the resulting solutions.  This requires checking
that the proper acceleration is small and taking the string
coupling to be small so that the DBI action is a good
approximation to the effective action of the probe.  It further requires
determining the range of parameters and times for which the energy
in the probe does not back react significantly on the geometry, so
that the probe approximation remains valid. Finally, we must
also check that fluctuations about the solution, density
perturbations, and particle and string production are not too
large.  In Section 4 we will consider generalizations of this
system to include coupling to four dimensional gravity and other
sectors, including effects of a cutoff throat in the IR; we will
take into account new contributions to the action at the level of
its most relevant terms at low energy.

\para
In this section we start by writing down the effective action
applicable to the pure conformal field theory (CFT) without
gravity (infinite $AdS$ space). The DBI action for a probe
D3-brane
moving in
AdS${}_5\times$S${}^5$ can be found in \cite{bigreview} and we
will use their conventions. We concentrate only on radial
fluctuations, and set the field strength on the brane to vanish.
Working in field theory variables $\phi\equiv r/\alpha^\prime$, we
have
\be S=-\frac{1}{g^2_{YM}}\int\,dt\,d^3x\,
f(\phi)^{-1}\left[-\det\left(\eta_{\mu\nu}+f(\phi)\,\partial_\mu
\phi \,\partial_\nu\phi\right)^{1/2}- 1\right] \label{dbi}\ee
where for now we take the background brane metric to be flat
$\eta_{\mu\nu}={\rm diag}(-1,1,1,1)$,
and $f(\phi)$ is the harmonic function
\be f(\phi)=\frac{\lambda}{\phi^4}
\label{f}\ee
Expanding the action out in derivatives leads to a canonical
kinetic term for $\phi\equiv r/\alpha'$ and a series of higher order
derivative interactions. The potential cancels out, reflecting the
BPS nature of the D3-brane. In fact, for a single excited scalar
field $\phi$, this form of the higher derivative action can be shown
to hold even at weak coupling \cite{malda} without recourse to a
dual gravitational description: it is fixed by the
requirement of a non-linearly realised conformal invariance,
together with known non-renormalisation theorems.

\para
So in general the dynamics of $\phi$ in our strongly coupled CFT
differs in two important ways from a naive scalar field system
governed by \eqn{usualact}.  Firstly, the kinetic term is
corrected to that in \eqn{dbi}; as discussed above we will find
that the proper velocity approaches the speed of light so that the
quantity in the square root approaches zero, resulting in a system far
away from the regime where the two derivative action suffices.
Secondly, there is no potential $V(\phi)$. In later sections we
shall remedy the latter issue by generalizing to systems in which
potentials are generated, both by anti-branes and by considering a
field theory coupled to other sectors. The speed limit will remain
a crucial ingredient in our generalizations.

\subsection{Dynamics}

\para
Let us now study more explicitly the approach of the ${\cal N}=4$
SYM theory to the origin of its Coulomb branch, using the action
\eqn{dbi}. Since we are interested only in the time dependence of
the solution, we shall ignore the spatial derivatives, leaving us
with the action,
\be S=-\frac{N}{\lambda^2}\int\ d^4x
\phi^4\left(\sqrt{1-\lambda\dot{\phi}^2/\phi^4}- 1\right)
\label{Sglobal}\ee
%
To determine the late time behaviour\footnote{Readers who prefer
$u=1$ units may make the substitution in all expressions of this
type.} of this system, we first compute the conserved energy
density $E$, given by (up to a factor of $1/g^2_{YM}$)
\be
E=\frac{1}{\lambda}\,\phi^4\,\left(
\frac{1}{\sqrt{1-\lambda\dot{\phi}^2/\phi^4}}- 1\right)
\approx
\ft12\dot{\phi}^2+\ft18
\frac{\lambda\dot{\phi}^4}{\phi^4} +\ldots
\label{energy}\ee
where, in the second line, we have expanded around large $\phi$. The
first term is the canonical kinetic energy for a rolling scalar field,
with subsequent terms becoming important at small $\phi$, capturing the
effect of virtual W bosons.
\para
As mentioned above, the Born-Infeld action is valid for
arbitrarily high velocities $\dot{\phi}$ but only for small proper
accelaration $a$, \be a\sqrt{\alpha'}= \frac{\lambda^{1/4}}{\phi}
\frac{d}{dt}\left(\frac{\sqrt{\lambda}\,\dot{\phi}}{\phi^2}
\right) \nn\ee So in order for our analysis to be self-consistent,
we must check that the proper acceleration is much smaller than
string scale in our solution. Inverting \eqn{energy} to solve for
$\dot\phi$, and subsequently taking the time derivative, it is
simple to see that we can trust the DBI analysis when we are in
the strong coupling regime $\lambda\gg 1$.
\para
Integrating once, the trajectory of the scalar $\phi$ is given by,
\be
t-t_0= \frac{1}{\sqrt{E}}\int_{\phi_0}^\phi d\varphi\
\frac{1}{\varphi^2}\frac{\lambda E+\varphi^4}{\sqrt{\lambda
E+2\varphi^4}}
\nn\ee
We are interested in the dynamics of the scalar field after we hit
the critical point on moduli space defined by \eqn{first}. Using
the classical expectation $E\sim v_0^2$, this means we are
interested in the regime
\be \phi^4\ll \lambda E \label{goodregime}\ee
Here we find the late time behaviour, \be
\phi(t)\rightarrow\frac{\sqrt{\lambda}}{t} \label{latelate}\ee
saturating the speed limit \eqn{limit}. We therefore find that
from the perspective of the field theory observer, Xeno is
vindicated and the scalar field takes an infinite time to reach
the origin as advertised.

\subsubsection{Background Check}

The above calculation treated the $D3$-brane as a probe and is
valid only when the  back reaction can be neglected. Since the
brane travels at almost the speed of light at late times on the
gravity side, it carries a lot of energy and we must determine
the conditions under which its back reaction does not
destabilize the AdS background.\footnote{This constraint was
obtained via discussions with M. Fabinger.} The gravitational
field surrounding a highly boosted object in locally flat space is
given by the Aichelburg-Sexl metric \cite{sexl}, which in our case of
codimension 6 is given by
\be h_{--}\sim {{l_s^8g_s E_p}\over r_\perp^3}\delta(x^-)
\label{ASmetric}\ee
Here $E_p=E(R/r)^4$ is the proper energy density of the probe
(related to the Poincare observer's energy density $E$ by the
appropriate powers of the warp factor $r/R$), $r_\perp$ is the
distance from the probe in the transverse dimensions, and $x^-$ is
the light cone coordinate with respect to which the brane
trajectory is localized in locally flat coordinates. If we insist
that the corresponding curvature ${\cal R}_{10}\sim
h_{--}/r_\perp^2$ smeared over a string scale distance across the
brane be smaller than the ambient AdS curvature $1/R^2$ at a
distance $R$ from the probe, this gives the condition
\be E < {\phi^4\over {\lambda^{1/4}g_s}} \label{br1}\ee
If we impose this condition, we still obtain a window
\be
\lambda^{1/4}g_s E < \phi^4 < \lambda E
\nn\ee
in which our back reaction constraint \eqn{br1}\ intersects with
our regime \eqn{goodregime}\ of limiting speed.

\para
This constraint \eqn{br1}\ may be too strong, since the novel
behavior we found for $\phi$ based on the DBI action depends only
on the AdS geometry.  We can estimate the back reaction of our
probe on the AdS geometry as follows. The probe forms a domain
wall of energy density $E_p$ in the five dimensional gravity
theory obtained by dimensional reduction on the $S^5$.  Such a
wall will jump the warp factor accross it.  By dimensional
analysis, this gives the relation
\be {1\over R^\prime}-{1\over R} \sim E_p l_5^3 \label{warpjump}
\ee
in terms of the five dimensional Planck length $l_5$, where
$R^\prime$ is the curvature radius on the IR side of the domain
wall. Using $l_5^3\sim l_{10}^8/R^5=l_s^8g_s^2/R^5$, the condition
that the jump in warp factor is smaller than the original AdS warp
factor is
\be
E< {\phi^4\over g_s}
\nn\ee
Again, this leads to a window at strong coupling in which the
rolling scalar field saturates its speed limit \eqn{latelate} in
the regime
\be
g_s E <  \phi^4  < \lambda E
\label{br2}\ee
Similar back reaction criteria apply to all the cases discussed in
the paper, and we will derive a related bound in our interesting
inflationary phase.
It would be interesting to explore what happens when
this condition is violated.  We will discuss the other consistency
conditions for our background (including ruling out significant
back reaction from particle production) in Section 6.

\subsection{Antibranes}

Here we introduce the first of our generalizations: anti-D3-branes
moving in the AdS background \eqn{AdSmetric}. The action for the
$\overline{D3}$-brane probe in $AdS_5$ differs from the $D3$-brane
case \eqn{Sglobal} merely by a change of sign in the final term,
\be S=-\frac{N}{\lambda^2}\int\ d^4x
\phi^4\left(\sqrt{1-\lambda\dot{\phi}^2/\phi^4}+1\right)
\label{dbino}\ee Upon expanding the square-root in powers of
$\lambda\dot{\phi}^2/\phi^4$, this gives rise to a potential which
is quartic in $\phi$. There is no quadratic $m^2\phi^2$ term. This
may be at first sight surprising since the system has broken
supersymmetry in the presence of the antibrane, and loops of open
strings which probe the susy breaking are included at the level of
the action \eqn{dbino}.  However this result is to be expected
using simple Randall-Sundrum ideas.  The local string scale at the
position of the brane is $\phi/\lambda^{1/4}$ \eqn{MS}. Similarly
all hard masses are warped down by a factor of
$r/R=\phi\sqrt{\alpha'}/\lambda^{1/4}$ at position $\phi$ in the
throat. Applying this to the susy breaking scale $m_{SUSY}$, a
potential $V\sim m_{SUSY}^2\phi^2$ really scales like $\phi^4$.
Said differently, there is no independent scale in the CFT on the
Coulomb branch beyond the position $\phi$.
\para
The $\phi^4$ potential implies that the
$\overline{D3}$-brane experiences a force towards
the horizon at $r=0$. Nonetheless, it is a simple matter to check
that the late time dynamics is dominated by the speed limit and
is identical to that of the $D3$-brane \eqn{latelate}.

\para
In some sense, the $\overline{D3}$-brane probe in $AdS$ space
can be thought of as a strong
coupling limit of the usual tachyon matter system. Indeed,
we shall see in Section 5 that the resulting dynamics bears some
similarity  to the results that have been obtained for
the much-studied open string tachyon decay in the weakly coupled
brane-antibrane system (see \cite{ashoke} for recent review of
this system). However, there are some differences which we
stress here. At weak coupling, the $\overline{D3}$-brane system has
tachyons from strings stretching between the antibrane and the $N-1$ D3-branes
described
by the AdS throat.  However, at strong coupling $\lambda\gg 1$, it is easy to
see from \eqn{MW} and \eqn{MS} that this mode is not tachyonic.
The negative contribution to the would-be tachyon mass squared,
scaling like the (warped) string scale
$m_s(\langle\phi\rangle)^2$, is dwarfed by the positive
contribution $m_W^2$ to the mass squared from the stretching of
the string from the antibrane to the AdS horizon. Thus the
brane-antibrane system at strong 't Hooft coupling (obtained by
taking many branes and a single antibrane) has no tachyon.  The
system {\it is} unstable -- as we will discuss in detail the
antibrane falls towards the horizon. However this evolution is not
condensation of a brane-antibrane tachyon since none exists; as
in \cite{bfannihilation}
the annihilation of the antibrane against the unit of $D3$-brane
charge encoded in the RR flux is a nonperturbative effect.

\section{Coupling to Gravity}

In both the brane and antibrane cases, we will be interested in
two further important generalizations.  One is coupling to four
dimensional gravity, as well as to other sectors suppressed by
higher dimension operators that may arise in the corresponding
string compactifications.  The other is a generalization to a
throat which is capped off in the region corresponding to the IR
behavior of the field theory -- i.e. the system with a mass gap.
In a string theoretic setup, these effects can be achieved by
gluing an AdS-throat like solution onto a compactification
geometry which acts as a UV-brane as in \cite{H. Verlinde,GKP}.

\para
In this section we shall discuss the effective actions which
result from these generalizations, focusing on the relative
strength of corrections to the potential energy and DBI kinetic
energy terms in the effective action. Our motivation for
considering these theories is the application of the speed limit
mechanism to cosmology. We will postpone a full discussion of this
until Section 5.  To avoid undue suspense, let us here summarize
some of the important results.
\para
Because our matter sector itself slows the scalar field $\phi$
in its progress toward the origin (in the field theory
variables), the conditions for slow roll inflation are modified in
a useful way. In a situation with a strong enough additional
potential energy for $\phi$ (which may arise for example from a
setup with a brane or antibrane in a cutoff AdS throat coupled to
other sectors) we will indeed find a phase of slow roll inflation.
This can be obtained at subPlanckian VEV for $\phi$, with the slow
roll provided by the DBI kinetic term corrections rather than from
the usual gravitational damping or from an unnaturally flat
potential. This allows us to evade some of the problems discussed
in \cite{dflation}.  While we have not yet produced a complete
model of real-world inflation from this mechanism, it seems a very
promising ingredient.

\para
More generally, we will find some familiar behaviors for the FRW
scale factor $a(t)$ but arising in unusual ways from our matter
sector. For example, as in ``tachyon matter" we will find a dust
equation of state for motion of a D3-brane toward the origin of
the Coulomb branch, though in our case production of massive
matter is suppressed. With other potentials, we also find a novel
steady state late-time behaviour for the universe that does not
involve fine tuning of initial conditions.

\para
The motion of the moduli themselves is an important aspect of
string cosmology, and the basic result of the previous sections
leads to interesting novelties in this area.  In particular, if
the moduli field $\phi$ corresponding to the motion of the
threebrane heads toward the origin of the Coulomb branch, it gets
slowed down and, at least in the probe limit, takes an infinite
time to reach the origin. This is radically different from the
naive treatment based on the moduli space metric in which it would
shoot past the origin without pausing. Furthermore, if the scalar
relaxes into a potential well surrounding the origin, it does not
oscillate around the minimum  in the same manner as moduli treated
with the usual action \eqn{usualact}. However in the cases when
the $\phi$ sector with all the DBI corrections leads to a dust
equation of state, it can cause similar problems to the usual
moduli problem.

\subsection{CFT Coupled to Gravity}

We firstly ask what becomes of the low-energy effective action
\eqn{dbi} when
the gauge theory is coupled to gravity by introducing a dynamical
background metric $g_{\mu\nu}$. At the same time, we add a
potential term $V(\phi)$ to the action that may arise when the
system is coupled to four dimensional gravity and other sectors
involved in a full string compactification. A simple four
dimensional covariantization of \eqn{dbi} is
\be
{\cal L}_{0} =
-\frac{1}{g^2_{YM}}\,\sqrt{-g}\,\left( f(\phi)^{-1}
\sqrt{1+f(\phi)\,g^{\mu\nu}\partial_\mu\phi\,\partial_\nu\phi}
+V(\phi)\mp f(\phi)^{-1}\right)
\label{fullact}\ee
where the $\pm$ refers to the $D3$-brane and $\overline{D3}$-brane
respectively while $f(\phi)=\lambda/\phi^4$ as before.

\para
In the presence of four dimensional gravity, there are in general further corrections to the action
\eqn{fullact}\ coming from the following considerations. Dynamical $4d$ gravity leads generically to nontrivial
four dimensional curvature ${\cal R}$.
In effective field theory, this leads to series of contributions of the
form
\be {\cal R}\phi^2(1+c_1{\cal R}/\phi^2+\dots).
\label{curvseries}\ee
More generally each contribution in the action \eqn{fullact}\ could be corrected by terms suppressed by powers
of ${\cal R}/\phi^2$.  In particular, there will be corrections depending on ${\cal R}$ and $\partial\phi$,
which will be outside the scope of our analysis; our results will be contingent on these corrections being
subdominant (which may require tuning in appropriate circumstances). In the weakly coupled ${\cal N}=4$ SYM
theory coupled to four dimensional gravity, such corrections are evident from diagrams containing loops of W
bosons. At strong coupling, such corrections were found in \cite{sw,tsey}\ starting from the DBI action for
curved slices. From all these points of view, the coefficients $c_i$ in \eqn{curvseries}\ are expected to be of
order one (i.e. not parametrically large or small as a function of $\lambda$).

\para
In our solutions, we will find that these curvature corrections
are negligible. Our strategy will be to simply ignore them at this
stage, analyze the dynamics, and then check that this is self
consistent (in Section 5.3 below).  In any case, we can collect
these effects in an effective Lagrangian containing both gravity
and matter:
\be S=\int\,\left(\ft12\sqrt{-g}\,(M_p^2+\phi^2)\, {\cal R}+{\cal
L}_{0}+\dots\right) \label{whatiwant}\ee
where the $\dots$ refers to corrections to the terms in
\eqn{whatiwant}\ which are down by powers of ${\cal R}/\phi^2$
from the leading terms. The final action is a special case of
those considered in k-inflation and k-essence scenarios
\cite{kflation,kessence}. In our case, the higher derivative terms
in ${\cal L}_0$ follow from the strong coupling dynamics of the
field theory near the origin of its moduli space.

\para
Let us now discuss the reliability and plausibility of this action
including the coupling to four dimensional gravity.  In string
theory, such a coupling is obtained by embedding the AdS geometry
into a compactification as a throat emanating from a Calabi Yau
compactification of type IIB string theory \cite{H. Verlinde,GKP}.
The first question that arises is whether this geometrical
combination exists or if instead the coupling of the AdS to the
Calabi Yau produces a large deformation of the AdS space. If such
a geometry exists, we may move onto the second question: whether
the low energy effective field theory action is given by
\eqn{fullact}\ to a good approximation for the system we wish to
study, and what is the order of magnitude of the parameters in the
potential $V$. Thirdly, in the end we must check that the time
dependent solutions we find are stable against small fluctuations,
particle production, and density perturbations.  (This is in
addition to the checks for small acceleration and small back
reaction of the probe energy of the sort we completed in Section 3
for the global case.)  This last question will be addressed in
section 6.

\para
As far as the first question goes, as discussed in \cite{H.
Verlinde, GKP}, the AdS throat is obtained from a collection of
many D3-branes at a smooth point on the Calabi Yau.  At the level
of the no scale potential obtained for CY compactifications of
type IIB at large radius in the absence of $\alpha'$ corrections
and nonperturbative effects \cite{GKP}, the scalars on the
D3-branes have a potential identical to that in the global quantum
field theory on their worldvolume.  They are mutually BPS and, in
this sense, are also mutually BPS with the rest of the
compactification.  This follows from the analysis in \cite{GKP},
and may be related to the fact that in the ${\cal N}=4$ SYM at
strong coupling, the operator $tr\Phi^2$, which is relevant at
weak coupling, is highly irrelevant at strong coupling (something
exploited by e.g. \cite{Strassler}). The quantum mechanical
corrections to the Kahler potential and the superpotential of the
model, which are important for fixing the Kahler moduli, will in
general also determine the positions of the D3-branes.  Since the
configuration corresponding to their being on top of each other at
a smooth point in the Calabi Yau is an enhanced symmetry point, it
is likely that an order one fraction of the models have a minimum
containing to a good approximation an AdS throat arising from such
a collection of D3-branes.  We will momentarily estimate the size
of corrections to this geometry as seen by our probe.

\para
Now let us discuss the second question about the reliability of
our effective action and the parameters in the potential. Attached
to the Calabi Yau, our AdS throat is coupled to four dimensional
gravity and other sectors.  We will start by discussing the
effects of the coupling of these sectors to $\phi$ in effective
field theory. Then we will discuss the geometrical description of
these effects. We will argue that (consistently in both
descriptions) significant corrections to the mass term in the
potential can naturally arise from these couplings while
corrections to the kinetic terms arising from them are subleading
to those in the original action. This ensures that we can preserve
our slow roll effect for the scalar field coming from the kinetic
terms while introducing potential energy sources for gravity which
lead to interesting, and in one case accelerating, cosmologies. As
we will also discuss, it is an interesting open question whether
this setup requires tuning couplings of throat Kaluza Klein modes
to other sectors to avoid further effects on the kinetic terms for
$\phi$.

\subsubsection{Effective Field Theory Description of Corrections}

As well as curvature couplings such as
${\cal R}\phi^2$ arising from the introduction of gravity, we will
generically have further couplings of $\phi$ to other sectors. These could be of the form,
\be \phi^2\eta^2\ \ \ \ {\rm or}\ \ \ \ \
{{\phi^2(\partial\eta)^2}\over M_*^2} \label{gutcoupling}\ee
where $\eta$ is a field from another sector and $M_*$ is a mass
scale (such as the GUT, string, or Planck scale) in the system
above the energies we wish to consider. In the next subsection we
will explain why we expect such couplings to be available
naturally in string compactifications, but we have not constructed
an explicit example and will largely treat these couplings by
effective field theory in our analysis of the cosmology.

\para
We should emphasize that the effective field theory description we
use in this subsection makes use of the weak coupling expansion
available on the gravity side of AdS/CFT; it is just the couplings
of the $\phi$ fields to the new sectors involved in the
compactification which we will treat here via effective field
theory.  A full treatment of the 4d effective field theory on the
gravity side also requires understanding the couplings and effects
of the Kaluza Klein modes.  This involves understanding what
tuning is required to preserve an (approximate) AdS throat in a
Calabi Yau compactification in the absence of a probe brane.  This
is not well understood.  In any case, as we will discuss in the
next subsection and in Section 5, the probe approximation for the
brane in a large-radius approximate AdS throat is consistent with
the presence of a large enough mass on the probe brane for
interesting cosmological effects such as inflation.

\para
More generally, we may view \eqn{gutcoupling}\ as coupling the
strongly coupled CFT to the $\eta$ sector by hand in effective
field theory. The action \eqn{fullact} contains no closed string
(gravity) loops, or loops of $\eta$ from couplings such as
\eqn{gutcoupling}. We must estimate contributions to our effective
actions from these loop effects. We will assume a weak enough
string coupling for perturbative string theory to be valid on the
gravity side.

\para
Out first important task is to determine the scale of any relevant
or marginal corrections to the action that may arise from the
couplings involving $\phi$. Let us start by discussing the
potential $V(\phi)$, considering a power series expansion
\be V\equiv V_0 + V_2\phi^2 + V_4\phi^4 + \dots \label{Vexp}\ee
The most relevant term in the Lagrangian at low energies is the
hard cosmological constant $V_0$.  This gets contributions in
principle from all sectors in the system. In effective field
theory (and approximately in string theory using the Bousso
Polchinski mechanism \cite{BP}) we may tune this (close) to the
value of interest for a given application.  We will make use of
this freedom in our analysis.

\para
Now let us move on to the $\phi$-dependent couplings, starting
with the mass term $V_2\phi^2$. As we discussed in the global
case, in an exact conformal field theory (which is either not coupled
to gravity and other sectors or is not destabilized by their presence),
a mass coupling $m_{*}^2\phi^2$ with $m_*$ a constant mass
parameter is ruled out by the conformal invariance encoded in the
effects of the warping in the AdS throat. Thus, with an exact
$AdS$ throat geometry, the potential is expected to receive
corrections starting with the quartic $V_4\phi^4$ coupling.
However, in general the couplings to gravity and
other sectors can generate corrections to
the theory \eqn{whatiwant}\ which violate conformal invariance.
For example, corrections to the scalar mass coming from loops
containing virtual gravitons yield a contribution,
\be
m^2_{*}\sim \Lambda_{UV}^4/M_p^2
\label{mgrav}\ee
where the fourth power of the UV loop momentum cutoff
$\Lambda_{UV}$ can be traced to the fact that the Feynman diagrams
contain derivative couplings. Similar contributions arise from
couplings to other sectors \eqn{gutcoupling}. A mass for $\phi$ can
be generated either from a VEV for $\eta$, or from loop
corrections involving the derivative coupling, giving rise
to respective contributions of the form
%
%
%
%
%
%
%
\be
m_*^2\sim \langle\eta\rangle^2\ \ \ \ {\rm or}\ \ \ \ \
m_*^2\sim \Lambda_{UV}^4/M_*^2
\label{mgut}\ee
We should also note that the quartic and higher couplings
$V_4\phi^4+\dots$ in the effective potential will also get
corrections from interactions such as \eqn{gutcoupling}.  In our
strategy for obtaining inflation below, we will require the net
quartic and higher couplings to be small enough relative to the
quadratic coupling.  This will not require significant tuning of
parameters in our solutions.

\para

Now let us consider the strength of corrections to the crucial
generalized kinetic terms in the DBI action (i.e. the series in
$v_p^2=\lambda\dot\phi^2/\phi^4$) appearing in the action.
Importantly, the corrections to these terms coming from
\eqn{gutcoupling}\ are suppressed relative to the existing terms
that generate our slow roll effect (in contrast to the situation
for the soft mass term just described, for which the \eqn{mgut}
constitutes the leading effect.) For example, if we compute the
effect of the coupling \eqn{gutcoupling}\ on the $\dot\phi^4$ term
in the effective action via a loop of $\eta$s, we obtain a
contribution scaling like \be \left(\frac{\dot\phi}{M_*}\right)^4
\log(\Lambda_{UV}/M_*) \nn\ee This is much smaller than the
original contribution $N\dot{\phi}^4/\phi^4$ in the small $\phi$
regime of interest: $\phi^4\ll NM_{*}^4$.

\para
The fact that the DBI kinetic terms are robust against large
corrections from \eqn{gutcoupling}\ arises from the fact that they
are marginal and get logarithmic corrections suppressed by the
coupling constants in terms like \eqn{gutcoupling}. There are no
soft contributions to the derivative terms in the action--they
would have to be of the form $(m_*/\phi)^n(\dot\phi^2/\phi^4)$.
Such corrections would be more infrared divergent as $\phi\to 0$
than the existing ones in \eqn{dbi}.  There are no perturbative
diagrams producing such effects, and we find it implausible also
in the strong coupling regime that coupling to gravity could
worsen the IR behavior of the system.

\para
Although these corrections to the DBI kinetic terms are small,
they are nonzero in general.  Since the kinetic terms arose from
the DBI action given by the volume of the brane embedded in the
ambient geometry, corrections to them correspond to corrections to
the AdS geometry at least as seen by our probe brane. Indeed once
we generate (or add by hand in effective field theory) a
deformation of the theory by a mass term for $\phi$, we expect
deviations from exact AdS geometry in the IR.  However, as above
these deviations are small effects on our probe evolution relative
to the effects of the original kinetic terms. In particular, if we
deform the Lagrangian to introduce an $m^2\phi^2$ term, we can
estimate its effects on the higher derivative terms in the DBI
action.  These came from integrating out ``W bosons", with the
importance of the higher derivative terms arising from the fact
that these modes become light at the origin $\phi\to 0$.  The mass
term for $\phi$ will at some order induce new contributions to the
mass of the W bosons; up to logarithms these will at most scale
like
\be
m_W^2\to \phi^2 + g^2_{YM}m^2
\label{worstWmass}\ee
In this manner, the contributions of virtual W bosons to the DBI
Lagrangian for $\phi$ become a series in
$\lambda\dot\phi^2/(\phi^2+g^2_{YM}m^2)^2$.  This means that for $\phi\gg g^2_{YM}m$,
the
evolution is well approximated by that given by the original DBI
Lagrangian.  As we will see, this constraint preserves a window of
interesting behavior for $\phi$ dictated by,
\be
\phi^2>g_sm^2
\label{betterconstr}\ee
In Section 5.1, we shall see that the same constraint arises on the
gravity side by requiring small back reaction so that the
probe approximation holds.

\para
In summary, the same couplings which give rise to a mass $m$ for
$\phi$ do not badly alter the generalised kinetic terms for $\phi\gg m$.
In fact, the mass term itself is a small effect on the overall probe evolution
as we will see from the $\phi$ equation of motion in which the
DBI kinetic terms dominate. However, the mass term in the
potential will be an important source for the four dimensional
spacetime geometry in our cosmological solutions. The corrections
on the kinetic terms for $\phi < m$ do mean that
ultimately the geometry is probably
better approximated by a cut off throat. We will turn to this
analysis in subsection 4.2.

\para
As emphasized above, the four dimensional effective field theory however contains more than just $\phi$ and the
BPS W bosons, but also Kaluza Klein modes in the $AdS$.  It is important to check whether couplings similar to
\eqn{gutcoupling}\ and curvature couplings appear with $\phi$ replaced by Kaluza Klein modes.  If so, and if the
induced masses $m_{KK}$ of these Kaluza Klein modes were as big as $m$, the corresponding deformation of the
geometry in the IR region would remove our inflationary solution to be described in section 5.  In the context
of effective field theory, we may tune away these couplings if necessary, and as discussed in the next section
we generically expect such tuning to be possible in string theory. It would be interesting to determine whether
such tuning is necessary or if $m_{KK}\ll m$ appears naturally.

\subsubsection{Gravity Description of Corrections}

In the previous subsection, we gave arguments based on simple
effective field theory couplings that the (soft) mass term for
$\phi$ generated by \eqn{gutcoupling}\ will generically be
affected more strongly (relative to the original mass term for
$\phi$ in the Lagrangian) than the generalized kinetic terms
(relative to the original generalized kinetic terms in the
Lagrangian). At first sight, this may appear surprising from the
point of view of the geometrical picture of a brane probe.  Such a
probe has kinetic terms determined by the metric of the ambient
spacetime, and potential terms introduced via couplings to other
background fields (such as the five form RR field strength in the
$AdS_5\times S^5$ solution of type IIB supergravity). There is no
general relation between these two effects, though in particular
examples they are related in particular ways via the coupled
equations of motion for the metric curvature and the other low
energy fields. In familiar examples, such as the $AdS_5\times S^5$
solution, or the Klebanov Strassler solution \cite{ks}, the
strength of the potential is smaller than the effects we found in
the previous subsection.

\para
These classical solutions however do not include the effects
introduced by the other sectors (mocked up by $\eta$ in the above
analysis) located in the bulk of a Calabi Yau to which the throat
attaches.  In general, the relation between the spacetime metric
and the other background fields is different from that obtained in
the familiar classical solutions.  More general flux backgrounds
(including AdS solutions) with different types of brane probes
will have a different balance between fluxes felt by the probe and
metric curvature due to the contribution of other ingredients. Our
results in the above subsection reflect an aspect of this in the
context of a coupling of $\phi$ in the familiar solutions to other
sectors via, for example, an embedding in a Calabi Yau.

\para
To make this more concrete, let us use known scales of couplings
between different sectors of a Calabi Yau geometry to estimate the
effective field theory parameters $M_*$ and $\Lambda_{UV}$ in the
effective field theory analysis of the last subsection.  In
\cite{Dim} couplings between
brane throats were computed.  The results were consistent with
couplings between a brane throat and another sector (which could
be a much smaller throat) suppressed by powers of $M_*\sim 1/R$.
Taking the UV cutoff $\Lambda_{UV}$ of the second sector to also be of
order $1/R$, one obtains an estimate for the mass $m$ of order
$m\sim 1/ l_s\lambda^{1/4}$.  As discussed in the previous
section, the same coupling yields a suppressed correction to the
kinetic terms if we lie within the window
\be   {1\over {l_s\lambda^{1/4}}}  < \phi < {N^{1/4}\over
{l_s\lambda^{1/4}}} \label{octopus}\ee
To summarise: the important corrections to the mass term for
$\phi$ we derived via effective field theory couplings in the last
subsection fit with known properties of geometrical embedding of
brane probes, taking into account effects generated by sectors
outside the brane throat corresponding to our original CFT.
Therefore we find it likely that such a mass term is available
(and potentially generic given the presence of a large brane
throat) in explicit string models.  In any case, a mass for $\phi$
is physically consistent with a probe approximation for $\phi$.
This follows from the above considerations regarding the many
independent ways of assembling ingredients such as the target
space metric and other background fields, combined with the self
consistency against back reaction (to be checked in section 5) of
the brane energy carried in the $\phi$ mass term.

\para
However, we have not constructed an explicit example of this in a
full compactification model. It is a subtle problem to determine
the couplings among sectors in a full string compactification
\cite{Dim}\cite{dflation}, and we plan to pursue it systematically
in future work.


\subsection{QFT with Mass Gap Coupled to Gravity}

Motivated by the discussion above, we here consider geometries dual to
theories with a mass gap, where the AdS throat is cut off at the IR
end. The prototypical example of such a geometry is the
Klebanov-Strassler solution of \cite{ks}. The effects of such an
IR cut-off include the following.  The warp factor $f(\phi)$ in
\eqn{f} is changed, and further corrections to the action may
arise if we start from a cutoff throat at the classical level.
Here we shall consider only  a toy model in which we focus on the
effect on $f(\phi)$; similar comments to those in the previous
subsection apply to the question of further corrections to the
action.


\para
For the purposes of our discussion, we replace the full Klebanov-Strassler solution with a simple toy model
which reproduces the relevant features. We motivate this by considering a theory in which the formula for the W
boson mass \eqn{MW} is deformed to \be m_W^2=\phi^2+\mu^2 \label{wmass}\ee where, in the effective field theory
discussion above, we had $\mu^2\sim g^2_{YM}m^2$. In general $\mu$ will be related to $m$ in a similar fashion,
but for the purposes of this section we keep it arbitrary. To model the W mass \eqn{wmass}
 from the holographic perspective, we keep the action
\eqn{fullact}, but replace the warp factor $f$ with the
appropriate function. For simplicity we shall consider, \be
f(\phi)=\frac{\lambda}{(\phi^2+\mu^2)^2} \label{toy}\ee Such a
geometry does not satisfy Einstein's equations by itself, but
nonetheless exhibits the important features of a capped off
throat.
\para
In the DBI Lagrangian \eqn{fullact} for the cutoff throat
\eqn{toy} in the antibrane case, there is a hard mass of order
\be
m^2_{DBI}\sim \mu^2/\lambda
\label{mdbi}\ee
Loop corrections
arising from the $\phi^4$ interactions drive this up to
$m^2_{DBI}\sim \Lambda^2_{\phi}/\lambda$ where $\Lambda_{\phi}$ is
the ultra-violet cut-off pertaining to the $\phi$ loop.
\para
In our cutoff geometry we also have corrections to the masses and
other couplings as discussed in the last subsection (where now
there may be some new sectors associated with the physics at scale
$\mu$).  Again the effects on the kinetic terms are subleading but
the mass corrections are important.

\section{Cosmological Solutions}

In Section 3, we saw that the scalar field motion in our system is radically different from the naive
expectation based on the finite distance to the origin of the moduli space in the metric appearing in the two
derivative action \eqn{usualact}.  Here we shall ask what effects this novel matter sector has on cosmological
solutions. We will investigate this in detail by studying the possible FRW cosmologies solutions which follow
from \eqn{fullact}. We consider only spatially flat cosmologies, \be ds^2&=&-dt^2+a(t)^2dx^2 \nn\ee Since
spatially inhomogeneous terms are redshifted away during inflation, we consider the scalar field ansatz
$\phi=\phi(t)$. With this ansatz, the equations of motion can be concisely expressed by first defining the
analog of the Lorentz contraction factor in special relativity, \be \gamma \equiv
\frac{1}{\sqrt{1-f(\phi)\dot{\phi}^2}} \label{defgamma}\ee Our stategy is to analyze the solutions following
from \eqn{fullact} by first ignoring the effects of the curvature coupling ${\cal R}\phi^2$, and later showing
that its effects are self consistently negligible on our solutions (unlike the situation in usual inflationary
models based on \eqn{usualact}). The energy density $\rho$ and pressure $p$ following from \eqn{fullact} (taking
the $+$ sign, corresponding to a D3-brane, for definiteness) are given by,
\be
\rho&=&\frac{\gamma}{f}+(V-f^{-1}) \label{rho}\\
p&=&-\frac{1}{f\gamma}-(V-f^{-1})
\label{p}\ee
These
definitions do not include the overall coefficient of $1/g^2_{YM}$
in \eqn{fullact} which instead combines with $M_p$ in the Einstein
equations so that the scale $(M_p\sqrt{g_s})=M_pg_{YM}$ appears in
all equations. The Friedmann equations read,
\be
3H^2&=&\frac{1}{g_sM_p^2}\ \rho \label{f1}\\
2\frac{\ddot{a}}{a}+H^2&=&-\frac{1}{g_sM^2_p}\ p \label{f2} \ee
where $H=\dot{a}/a$ is the usual Hubble parameter. Finally, the
equation of motion for $\phi$ reads, \be
\ddot{\phi}+\frac{3f^\prime}{2f}\dot{\phi}^2-
\frac{f^\prime}{f^2}+\frac{3H}{\gamma^2}\dot{\phi}
+\left(V^\prime+\frac{f^\prime}{f^2}\right)\frac{1}{\gamma^3}=0
\label{fphi}\ee where prime denotes a derivative with respect to
$\phi$. Note that at as we approach the speed limit on moduli
space, so $\gamma\rightarrow \infty$, both the friction term and
the potential term in the equation of motion for $\phi$ become
subdominant. It is a simple matter to check that the second
Friedmann equation \eqn{f2} follows from the first \eqn{f1}
together with the equation of motion \eqn{fphi}. The effective
equation of state of our system, defined by $p=\omega\rho$, is
therefore given by \be
\omega=\frac{-\gamma^{-2}-(Vf-1)\gamma^{-1}}{1+(Vf-1)\gamma^{-1}}
\label{eqnstate}\ee

\subsubsection{The Hamilton-Jacobi Formalism}

As occurred in the global case, we will find solutions in which
the brane asymptotes to the speed of light at late times in the
gravity side background.  This means that the quantity
$\gamma$ grows as $t$ gets large, and the behavior
of $\rho$ will be substantially different from the usual case of
$\rho={1\over 2}\dot\phi^2+V$ to which it reduces for small proper
velocity.

\para
Firstly, we want to re-write the Friedmann equations \eqn{f1} and
\eqn{fphi} in a more tractable form. In fact, they can be
integrated once. In the inflation literature this is referred to
as the ``Hamilton-Jacobi'' formalism \cite{will}. Another perspective
can be obtained by viewing the
resulting cosmology as a Wick rotation of a BPS domain wall; the
first order Friedmann equations are related to the Bogomoln'yi
equations derived in \cite{dan}.

\para
To derive the Hamilton-Jacobi equations the important step is to
view the scalar field $\phi$ as the time variable. In practice,
this means we consider $H=H(\phi)$ with $\phi=\phi(t)$. This
immediately puts a limitation on the dynamics since it assumes
$\phi$ is monotonic and $H$ is a single valued function of $\phi$
and therefore does not allow for oscillatory behaviour in $\phi$.
We start by taking the time derivative of \eqn{f1} and, using
\eqn{fphi} along the way, we find
\be
6 H H^\prime \dot{\phi} =
-\frac{1}{g_sM^2_p}\,{3H\gamma\dot{\phi}^2}
\nn\ee
which can clearly be solved simply by
\be
\dot{\phi}=-2(g_sM^2_p)\, \frac{H^\prime}{\gamma}
\nn\ee
Since
$(\gamma^{-1})$ depends on $\dot{\phi}$, it's useful to invert
this to give us \be
\dot{\phi}=\frac{-2H^\prime}{\sqrt{1/(g_s^2M^4_p)+4f
H^{\prime\,2}}} \label{bog1}\ee Finally, we can substitute this
expression for $\dot{\phi}$ into \eqn{f1}, using the full
expression for $\rho$ from \eqn{rho} to get an expression for the
potential $V(\phi)$ in terms of the Hubble parameter $H(\phi)$,
\be
V=3(g_sM^2_p)\,H^2-\frac{(g_sM^2_p)}{f}\sqrt{1/(g_s^2M^4_p)+4fH^{\prime\,2}}+\frac{1}{f}
\label{bog2}\ee The advantage of these equations is that they can
be solved sequentially. Given a potential $V$, we solve \eqn{bog2}
to find $H(\phi)$. This can then be plugged into \eqn{bog1} to
find $\phi(t)$. Finally, this can be substituted back into
$H(\phi)$ to find the dynamics of the universe. We shall now solve
for the late time behaviour of these equations, both in the AdS
throat where $f=\lambda/\phi^4$, and in our toy model for the
cut-off geometry with $f=\lambda/(\phi^2+\mu^2)^2$.

\subsection{Cosmology in the AdS Throat}

We start by considering cosmology in the CFT, corresponding to a
brane moving in the full AdS throat. In this case, the first order
Friedmann equations \eqn{bog1} and \eqn{bog2} read
\be
\dot{\phi}=\frac{-2H^\prime}{\sqrt{1/(g_s^2M^4_p)+4\lambda
H^{\prime\,2}/\phi^4}}
\label{newbog1}\ee
and
\be
\lambda V=3\lambda
(g_sM^2_p)\,H^2-(g_sM^2_p)\phi^4\sqrt{1/(g_s^2M^4_p)+4\lambda
H^{\prime\,2}/\phi^4}+ \phi^4
\label{newbog2}\ee
We will consider potentials of the form
\be
V=V_0+V_2\phi^2+V_4\phi^4+\ldots
\label{whereditgo}\ee
and determine the behaviour of the scale factor as
$\phi\rightarrow 0$. For now, we will simply introduce these
parameters by hand to analyze the equations but, as mentioned
above, there are constraints on these coefficients in the
${\cal N}=4$ SYM and in this theory coupled to gravity which we should
include in the final analysis.
For now
however, we work with the ad hoc potential above. There there are
three possible cases, depending on which of $V_0$, $V_2$ or $V_4$
is the first non-vanishing coefficient. (We shall in fact see that
if the first two vanish, and the potential is positive, then
the result is universal).

\subsubsection*{Case A: $V_0\neq 0$}

If $V$ is simply a constant then equation \eqn{newbog2}
is solved by $H=\sqrt{V_0/3(g_sM^2_p)}$, in which case
equation \eqn{newbog1} shows that $\phi={\rm const}$, and we get a
standard de Sitter phase. Of course, our kinetic terms aren't
playing a role here since $\phi$ isn't moving. Note that equation
\eqn{newbog1} requires $V_0>0$ here.

\para
Suppose instead that we take $V=V_0+V_4\,\phi^4$ which suffices to
get $\phi$ moving. Then for small $\phi$, we may choose the simple
ansatz $H=h_0+h_4\phi^4+\ldots$ where, from \eqn{newbog2} we must
choose the coefficients  $h_0=\sqrt{V_0/3(g_sM^2_p)}$ and
$h_4=V_4/6(g_sM^2_p)\,h_0$. Substituting this into the first
of the Friedmann equations, we have the late time behaviour for
$\phi$,
\be
\phi\rightarrow
\frac{1}{4(M_p\sqrt{g_s})\sqrt{h_4}}\,\frac{1}{\sqrt{t}}
\nn\ee
We see that, as in the case of the global CFT considered in Section
3, the scalar field takes infinite time to reach the origin.
However, in this case the friction from the expanding universe
means that $\phi$ does not saturate the speed limit \eqn{limit}.
From the late time behaviour of $\phi$, we may immediately
determine the late time behaviour of the scale factor. We have,
\be
H=\sqrt{\frac{V_0}{3(g_sM^2_p)}}\left(1 +
\frac{3}{2^7V_4(g_sM^2_p)}\,\frac{1}{t^2}\right)
\nn\ee
This result highlights the peculiar cosmologies that can arise due to strong
coupling effects of the gauge theory. Here we find a standard de
Sitter phase, superposed with a cosmology in which the scale
factor has the form $a(t)\sim exp(-c/tM_p)$. We shall see more cosmologies
with this characteristic exponential behaviour in Case D.

\subsubsection*{Case B: $V_0=0$}

Let us now consider the interesting case that will give rise to
inflation. We suppose that $V_2\neq 0$ so that this quadratic mass
term dominates at late times when $\phi$ is small.

\para
We work with the simple ansatz,
\be
H=h_1\phi+\ldots
\label{h1}\ee
Substituting this into \eqn{newbog2},
we get contributions to $V_2$ from both the $H^2$ term as well the
$\sqrt{\cdots}$ term. We find the potential takes the form,
\be
V=\left(3h_1^2-\frac{2h_1}{\sqrt{\lambda}}\right)(g_sM^2_p)\,\phi^2+{\cal
O}(\phi^4)\equiv V_2\phi^2+ {\cal O}(\phi^4)
\label{oldun}\ee
which, solving for $h_1$ in terms of $V_2$ gives
\be
h_1=\frac{1}{3\sqrt{\lambda}}\left(1+\sqrt{1+3V_2\lambda/(g_sM^2_p)}\right)
\label{h1really}\ee
Substituting our ansatz for $H$ into the
equation \eqn{newbog1}, we find
\be
\dot{\phi}=\frac{-2h_1\phi^2}{\sqrt{\phi^4/(g_s^2M^4_p)+4\lambda
h_1^2}}
\nn\ee
As $\phi\rightarrow 0$, we simply throw away the
$\phi^4$ term in the denominator to find the late time behaviour
\be
\phi\rightarrow\frac{\sqrt{\lambda}}{t}+\ldots
\nn\ee
This
coincides with the result \eqn{latelate} from the global analysis
of Section 3; in this theory the slowing down of the scalar
field is driven by the speed limit rather than friction from the
expanding spacetime. We now substitute this back into our ansatz
\eqn{h1} to find the late time behaviour of the scale factor,
\be
a(t)\rightarrow a_0\, t^{h_1\sqrt{\lambda}}
\label{growing}\ee
Using \eqn{h1really}, the exponent is
\be
h_1\sqrt{\lambda}=\frac{1}{3} \left(
1+\sqrt{1+\frac{3V_2\lambda}{(g_sM^2_p)}}\right) \approx
\sqrt{\frac{V_2\lambda}{3(g_sM^2_p)}}
\label{noreally}\ee
where the approximation is true only for $V_2\lambda\gg
g_sM^2_p$. We see that we can orchestrate an inflationary
universe if the potential $V_2$ is sufficiently large.
Specifically \be {{V_2\lambda}\over {g_sM_p^2}}
> 1 \label{bigbadv2}\ee It is amusing to recall that usual slow
roll inflation occurs only if the potential is suitably flat. Here
we find the opposite result: with a speed limit on the scalar
field, inflation occurs only if $V_2$ corresponds to a high enough
mass scale (we see from \eqn{bigbadv2}\ that this mass scale can
be much lower than $M_p$, but the acceleration gets stronger for
larger $V_2$).

\para
Interestingly, we have obtained a form of power law inflation, since the scale
factor behaves as a power of $t$.  However it arises in the case
of a polynomial (quadratic) potential, which under the usual
circumstances \eqn{usualact} leads to exponential inflation $a\sim
e^{Ht}$. (The potential required to obtain power law inflation
from \eqn{usualact}\ is exponential in $\phi$.)

\para
Naively it appears that this result suggests an alternative to the
usual quintessence scenario where a scalar field rolls towards an
asymptotic regime of moduli space. Here the scalar field could
roll a finite distance in the field theory moduli space metric in
the interior of the moduli space and still lead to asymptotic
acceleration. However, we are limited to a finite period of
controlled speed of light evolution because of the back reaction
of the probe on the geometry. In the global case we determined this
backreaction in \eqn{br2}. It is a simple matter to repeat the
calculation here where the energy of the brane is dominated by the
potential $V_2\phi^2$. Once again, translating from the proper energy
$E_p$ to the observers energy density $E=E_p(r/R)^4$, we have
the constraint
\be
E_p l_5^3\sim\frac{V_2\phi^2}{g^2_{YM}}\left(\frac{R}{r}\right)^4
\frac{\alpha^{\prime\,4}g_s^2}{R^5} < \frac{1}{R}
\nn\ee
which gives us the constraint that our solution is only to be
trusted for $\phi$ greater than
\be
\phi^2 > g_sV_2
\nn\ee
Notice that this is identical to the lower bound on $\phi$ arising
derived in Section 4 by calculating one-loop corrections to the
W-boson mass using effective field theory methods
\eqn{betterconstr}. Despite this infra-red cut-off to our inflationary
phase, we shall shortly argue that we still retain a viable window
of inflation in this scenario.

\subsubsection*{Case C: $V_0=0$ and $V_2=0$}

Let us now suppose that the potential
has only $V_4\neq 0$. As we have seen, such a potential is generated by
a $\overline{D3}$-brane moving in the AdS throat. The late-time analysis of the
Hamilton-Jacobi equations \eqn{newbog1} and \eqn{newbog2} is actually
just a special case of the above discussion, where we put $h_1=2/3\sqrt{\lambda}$.
We can read of the final result from \eqn{growing}: we have
\be
a\rightarrow a_0\,t^{2/3}
\nn\ee
which looks like dust. In fact, for any potential
for which $V_0=V_2=0$ with higher order terms non-vanishing, the speed
limit on the moduli space ensures that the rolling scalar field looks
has equation of state $\omega\approx 0$: a kinetic dust.

\subsubsection*{Case D: $V<0$}

To find equations of state with $\omega >0$, we require negative
$V_2$. From \eqn{h1really}, we see that for
$0<h_1<2/3\sqrt{\lambda}$ we have
\be
-\lambda(g_sM^2)^2\,\frac{\lambda}{3} < V_2 < 0
\nn\ee
The late time behaviour of the scale factor is
once again given by \eqn{growing}. During this expansion, the
scalar field is rolling up the inverted harmonic oscillator
potential. Without the higher kinetic terms, the scalar would
overshoot the top and roll down the other side. However, the speed
limit does not allow this to happen, and slows the field to
prevent it reaching the top at $\phi=0$ in finite time.

\para
Let us now consider the case $V_2=0$ with $V_4<0$. We can generate
such a potential by considering, for example, $H=h_3\phi^3$ which
gives rise to the potential $V=V_4\phi^4+V_6\phi^6$ where \be
\lambda V_4=-\sqrt{1+36\lambda h_3^2\,(g_s^2M_p^4)}+1\ \ \ \ ,\ \
\ \ \ \lambda V_6=3\lambda h_3^2\,(g_sM_p^2) \nn\ee At late times
the scalar field rolls like $\phi\sim 1/t$, and the equation of
state parameter diverges: $\omega\rightarrow\infty$. Meanwhile,
the scale factor $a$ has the peculiar behaviour \be
a=a_0\exp\left(-\frac{(1/(g_sM_p^2)+36\lambda
h_3^2)^{3/2}}{2^4\,3^3\,h_3^2} \,\frac{1}{t^2}\right) \nn\ee which
undergoes a period of accelerated expansion before settling down
to a steady-state cosmology at late times. The period of
acceleration coincides with the period when the $V_6\phi^6$ term
dominates over the $V_4\phi^4$ term in the potential; in the
solution this also correlates with a period in which the potential
dominates in the gravitational dynamics determined by $\omega$
\eqn{eqnstate}. As in case B, the effect becomes stronger with a
stronger potential (so a flat potential is not required for slow
roll since this is provided by the DBI kinetic terms).

\para
Scale factors with the characteristic $\exp(-1/t^2)$ behaviour
also occur in standard cosmology, but only for very finely tuned
inital conditions. The novelty here is that such dynamics is
generic for a large enough intitial velocity. A similar steady
state dynamics also arises from $H=h_4\phi^4$, this time with
$a\sim \exp(1/M_p t)$. Each of these scenarios finds Xeno
exonerated and Hoyle happy.

\subsection{Cosmology in the Cut-Off Throat}

The interesting Case B of the previous section has $V_2\neq 0$
which is not compatible with the conformal symmetry implied by an
exact AdS throat. Instead, one ultimately expects a geometry in
which the throat is cut off in the infra-red. We described such
geometries in Section 4. Here we will work with the toy model in
which the warp factor is taken to be \be
f(\phi)=\frac{\lambda}{(\phi^2+\mu^2)^2} \nn\ee For $\phi\gg\mu$,
the geometry is essentially that of the AdS throat and the
analysis of the previous sections holds. However, for $\phi\leq
\mu$, the speed limit becomes simply \be \dot{\phi}\leq
\frac{\mu^2}{\sqrt{\lambda}} \nn\ee So the scalar field can now
reach the origin in finite time and happily sail right through at
finite speed. This has the advantage that at late times the field
oscillates around the minimum of the potential as usual and
reheating may occur (depending on the couplings to the Standard
Model). However, we must check that we can still arrange for a
period of acceleration. The condition for acceleration requires
large enough $V_2$ as we saw in \eqn{bigbadv2}
Generally we have \be V_2=m^2 \nn\ee where masses
$m^2$ which could be generated in various ways were listed in
Section 4 and include contributions from graviton loop \eqn{mgrav},
couplings to other sectors \eqn{mgut} and a hard mass from the
deformed geometry \eqn{mdbi}.

\para
Let us start by examining the upper limit for $\phi$. In fact, we are quite entitled to take $\phi$ as large as
we wish and, for $\phi\gg M_p$, one finds the standard slow roll expansion of chaotic inflation. However, as we
shall discuss in Section 5.3, models with superPlanckian VEVs suffer from destabilization from a slew of quantum
corrections involving for example gravitational curvature couplings ${\cal R}\phi^2$ \cite{dflation}. To avoid
this, we shall instead restrict ourselves to {\it subPlanckian} VEVs
\be \phi \ll  M_p, \label{subplanck}\ee
where, as the speed limit is saturated, we may still obtain inflationary behaviour.\footnote{In our subsequent
analysis of density perturbations \cite{DBISky}, we discovered that to satisfy all the observational constraints
including those on non-Gaussianity, we require starting inflation at Planck scale $\phi$, which may require fine
tuning.}
\para
In our cut-off geometry we also need $\phi$ to be suitably large
so that the probe brane is experiencing the $AdS$ throat-like
background before the cut-off is reached. It can be checked that
the solution described in Case B of Section 5.1 continues to hold
in the cut-off geometry provided $\phi\gg\mu$. We therefore find
our window of inflation when the scalar field lives in the regime,
\be \mu\ll\phi\ll M_p \label{endpoints}\ee during which time we
have the cosmological solution $a(t)=a_0t^{h_1\sqrt{\lambda}}$
while the scalar field slows down as $\phi=\sqrt{\lambda}/t$. Let
us assume that this solution actually holds all the way to the end
points \eqn{endpoints}. This may be overly optimistic, but it
gives us a quick and ready way to calculate the number of
e-foldings. We require that at the initial time $t=t_0$ we have
$\phi(t_0=0)=M_p$, while at the final time $t=t_f$ we have
$\phi(t=t_f)=\mu$. Then, \be t_0={\sqrt{\lambda}\over M_p} \ \ \ \
{\rm and}\ \ \ \ \ t_f=\frac{\sqrt{\lambda}}{\mu} \nn\ee The
number of e-foldings is therefore given by, \be
n=\log\left(\frac{a(t=t_f)}{a(t=0)}\right)=h_1\sqrt{\lambda}
\log\left(M_p/\mu \right) \nn\ee The argument of the logarithm is
roughly the distance travelled by the scalar field. Any large
number of efoldings comes from the prefactor. Using \eqn{noreally}
we have \be n\sim
h_1\sqrt{\lambda}\sim\frac{m\sqrt{\lambda}}{\sqrt{g_s}M_p}\geq 100
\label{efoldings}\ee This criterion is the same as the criterion
that the potential $V$ dominate in the equation of state (in
particular in $\rho$).

\para
This inflationary phase is novel in several ways. Most importantly, the inflation can occur on a steeper
potential hill than works for ordinary slow roll inflation.  Moreover, at least before considering observational
constraints from density perturbations, we find inflation to occur at sub-Planckian field strength for $\phi$.
We will show in section 5.3 that, when combined with the novel form of the equation of motion for $\phi$, this
allows us to avoid destabilizing effects from a curvature coupling ${\cal R}\phi^2$, circumventing some of the
difficulties involved in placing inflation within a string compactification \cite{dflation}. Moreover, as noted
in Section 5.1, we have obtained {\it power-law} inflation from a quadratic potential.

\para
At later times, the brane reaches the end of the cutoff throat and
the novelties of the DBI action wear off.  At this stage, we
expect to reduce to an ordinary matter or radiation dominated
phase, though we have not yet controlled the details of the exit
from our inflationary stage.

\para
The number of e-foldings \eqn{efoldings} can be made large if we
can generate a large mass term $m^2\phi^2$ for our scalar field.
In Section 4, we discussed possible mechanisms for generating such
a term and argued that corrections to our crucial higher derivative
terms were sub-leading. The simplest of the mass corrections
was the hard contributions \eqn{mdbi} from the DBI action
itself: $m_{DBI}\sim\mu/\sqrt{\lambda}$. Comparing with \eqn{efoldings}
we immediately see that in our strong coupling regime $\lambda\gg 1$
there is no inflationary window with such a mass: we need another
mechanism. As discussed at length in Section 4, such a mechanism
may arise from coupling to other sectors which
(depending on the strength of the couplings and
cutoff scales for the other sectors in the system) may produce a
large enough mass.  More simply, as discussed above we may
consider deforming the theory by a mass term for $\phi$ and note
that this induces only subdominant corrections to the DBI kinetic
terms. Again, it would be very interesting to determine the
magnitude of mass $m$ arising in specific string
compactifications.

\subsubsection*{A Comment on $V_4$ Terms}

In the above discussion, we concentrated on the possible mass
terms that could be generated, and the associated window of
applicability of the inflationary solution described in Case B of
Section 5.1. However, the solution itself is valid only if the
quadratic term in the potential dominates over the quartic term;
otherwise we obtain the dust-like evolution of Case C. We
therefore require,
\be V_4\phi^2<m^2 \Rightarrow \phi^2 < m^2/V_4 \label{V2dom}\ee
If $V_4$ is too big, then this constraint on $\phi$ would be much
stronger than \eqn{subplanck} and potentially ruin our window
of inflation. Let us therefore examine the possible $V_4$ couplings.

\para
In the DBI action, $V_4$ scales
like $1/\lambda$. However, quantum corrections change this. Let's
focus on the virtual effects of other sectors, since these were
successful in giving us an inflationary window. The derivative
coupling \eqn{gutcoupling}, leads to a contribution of order
\be \Delta V_4\sim \Lambda_{UV}^4/M_*^4 \label{V4gut}\ee
while at the same time inducing a mass $m_*\sim
\Lambda_{UV}^2/M_*$ \eqn{mgut}\ as we discussed earlier.  Plugging
this into \eqn{V2dom}\ leads to the constraint $\phi < M_*$ which
is stronger than \eqn{subplanck}.  At the level of our effective
field theory analysis (in terms of an unknown UV cutoff parameter
$\Lambda_{UV}$ and a coupling scale $1/M_*$), this is consistent
with our inflationary window for appropriate values of these
parameters.

%
%


\subsection{Curvature Corrections}

Let us now restore into our analysis the curvature coupling ${\cal R}\phi^2$ and the other curvature corrections
scaling as powers of ${\cal R}/\phi^2$. We will consider the regime of our solutions where $\phi\ll M_p$, and
show that the solutions are not destabilized by the addition of these curvature couplings (in other regimes more
fine tuning would be required \cite{DBISky}). We should note here that the {\it conformal} coupling required to
render the stress energy tensor traceless in the case of large $\gamma$ will be more complicated than a ${\cal
R}\phi^2$ coupling.

\para
Let us first consider the ${\cal R}/\phi^2$ corrections to the
terms in the Lagrangian \eqn{whatiwant} which scale like
$H^2/\phi^2$.  In our inflating solution, these terms are of the
order $m^2/M_p^2$. As long as this quantity is sufficiently
smaller than 1, then the correction terms are negligible compared
to the terms in \eqn{whatiwant}\ that we did include in the
initial analysis. In the noninflating solutions, $H\sim 1/t$ and
these corrections are suppressed by powers of $\lambda$ which
makes them very safe from affecting the results above.

\para
Now let us turn to the curvature coupling ${\cal R}\phi^2$. First, let us consider the Einstein equations for
the metric. The effect of the conformal coupling is to replace $M_p^2$ with $M_p^2+\phi^2$ in these equations,
changing the effective Planck mass. Since $\phi\ll M_p$, this does not change the solutions for $a(t)$
significantly. Similar comments apply to the $H^2\phi^2$ terms which plague F-term inflation in supergravity
models.

\para
Secondly, we must check whether the scalar field equation of motion is self consistently solved by the original
solution in the presence of a curvature coupling. The coupling $\xi{\cal R}\phi^2$ contributes to the equation
of motion for $\phi$, \eqn{fphi}, as
\be \frac{\lambda\ddot{\phi}-6\lambda\dot{\phi}^2/\phi+4\phi^3}
{(1-\lambda\dot{\phi}^2/\phi^4)^{3/2}}+\frac{3\lambda H\dot{\phi}}
{(1-\lambda\dot{\phi}^2/\phi^4)^{1/2}}+(V^\prime +2\xi{\cal
R}\phi-4\phi^3)=0 \label{fphiconf}\ee
Because $\gamma(t)^{-1}=\sqrt{1-\lambda\dot{\phi}^2/\phi^4}$ is
going to zero at late times, the terms involving $V$ and ${\cal
R}$ are subdominant in this regime. This is reflected in the fact
that, for Cases B and C of Section 5.1, the late time behaviour of
the scalar field saturates the speed limit and is independent of
the potential.

\para
In summary, the curvature couplings we have considered here do not change our results. This is in contrast to
the usual situation \eqn{usualact}, where for example the slow roll conditions for inflation in a polynomial
potential are impossible to satisfy in the presence of the coupling ${\cal R}\phi^2$.  This point was emphasized
for the case of 3-branes in warped throats in \cite{dflation}. Here see that near the origin, where the DBI
velocity corrections are crucial, the structure of the scalar equation of motion is dramatically different due
to the factors of $\gamma(t)$ in the denominator in \eqn{fphiconf}.

\subsection{Relation to other Works}

The behavior of our system, especially in Case C in Section 5.1,
bears a strong resemblance to the tachyon matter system studied in
recent years \cite{tachyon,tacycosm}. Indeed, the anti-brane
moving in the $AdS$ throat is, in some sense, a strong coupling
limit of the tachyon matter system. There are important
differences between our case and the weakly coupled tachyon matter
system, most notably the fact that as we discussed above in our
situation the tachyon has been lifted and, as we shall see,
particle production is suppressed. Moreover, the spectra of
particles and strings in the two backgrounds differ.

\para
A simple field redefinition can be used to relate
the effective field theories in the two cases as
follows. In tachyon cosmology, the action for the tachyon is of
the form \cite{Garousi:2000tr}
\be
{\cal L}=-F(T)\sqrt{1-\dot{T}^2}
\label{tachlag}\ee
which, upon
expanding the square-root, has a potential term $F(T)$. We can
simply generalise this action to include a further potential term
$G(T)$, which we could use either to cancel $F$ or simply to make
the potential different from the coefficient of the kinetic terms.
Such a potential may be generated in situations where the brane
antibrane system is embedded in a compactification.  We therefore
consider, \be {\cal L}=-F(T)\sqrt{1-\dot{T}^2}-G(T) \nn\ee The
stress-energy tensor and pressure can be easily calculated to
yield energy density
\be
\rho=\frac{F}{\sqrt{1-\dot{T}^2}}+G\ \ \
\ ,\ \ \ \ p=-F\sqrt{1-\dot{T}^2}-G
\nn\ee
Tachyon cosmologists work with this Lagrangian imposing $G=0$. Of
particular relevance for the present work is the observation \cite{others}
that the power-law inflationary phase of Case B can be obtained by the
choice $F(T)\sim 1/T^2$ in the Lagrangian \eqn{tachlag}. It was further shown
that higher order (inverse) potentials give rise to
dust behaviour as we saw in Case C.
To return to our favourite Lagrangian and compare these results with those
above, we employ the field redefinitions
\be
T=\frac{\sqrt{\lambda}}{\phi}\ \ \ ,\ \ \
F(T)=\frac{\lambda^2}{T^4}\ \ \ \ ,\ \ \ \ G(T)=V(\phi)-\phi^4
\nn\ee
It is very interesting to ask how generally motion on
internal scalar field configuration spaces can be imbued with a
geometric interpretation (and therefore a speed limit induced by
causality). It is intriguing that the tachyon Lagrangian
\eqn{tachlag}\ has a structure reminiscent that of a relativistic
particle moving in spacetime.

\para
The motion of the probe brane in the cut-off geometry is also
somewhat similar to the discussion of \cite{bounce}. These authors
consider the induced mirage cosmology \cite{mirage} on a probe
brane as it moves in the Klebanov-Strassler geometry \cite{ks}.
This differs from our analysis in two important respects: firstly,
in \cite{bounce} the probe brane was taken to move slowly through
the background geometry, so that the DBI action could be
approximated by two-derivative terms. Secondly, the cosmology was
viewed from an observer on the probe brane, rather than from the
perspective of a boundary observer as is relevant for our
discussion.

\section{Perturbations and Particle Production}

We now come to the promised self consistency checks of our time
dependent evolution.   In order to trust our results in the
previous sections, we must check that particle and string
production and fluctuations about the solution do not destabilize
it.  We have already checked that acceleration is small enough to
avoid closed string production, and that back reaction from the
probe's large kinetic energy is small in the window \eqn{br2}.

\subsection{W boson and String Production}

Firstly, we consider the production of massive strings on the
brane and W-bosons. Our effective DBI action is blind to the
possible on-shell creation of these modes and we must check by
hand that the production is suppressed in the $\lambda\gg 1$
regime. This is rather simple. The time dependent solution we
found entails time dependent masses for both W-bosons \eqn{MW} and
massive string modes on the brane \eqn{MS} since these masses are
proportional to $\phi\sim{\sqrt{\lambda}/t}$.

\para
The strength of particle production for a time dependent frequency
$\omega(t)$ is controlled by ${\dot \omega/\omega^2}$. This is a
rough estimate which can be obtained by starting from the
requirement for adiabatic evolution for a frequency which changes
over a time period $\delta t$:
\be \delta t \gg 1/\omega_{min} \label{adiabatic}\ee
where $\omega_{min}$ is the smallest value of $\omega$ obtained in
the evolution. Dividing by the change $\delta \omega$ arising in the process
yields
\be \delta t/\delta \omega \gg 1/(\omega_{min}\delta\omega)
\label{nextcondad}\ee
Consider a long enough evolution so that $\delta\omega$ is of
order $\omega_{max}$.   Replacing $\delta\omega$ on the right hand
side by $\omega_{max}$ and replacing $\omega_{max}\omega_{min}$ by
$\omega^2$  yields a condition under which particle production may
be consistently neglected:
\be \dot\omega \ll \omega^2 \label{finalcondpp}\ee
This can be thought of as expressing the fact that (on average) the
energy from the time dependence of the frequency is smaller than
the jump in energy needed to produce an on shell particle.
We find in our situation that the inequality
\eqn{finalcondpp} is satisfied parametrically in
$\lambda$ for large $\lambda$.
In particular for the $W$ boson zero modes we find
\be {\dot m_W\over m_W^2}\sim {1\over\sqrt{\lambda}}
\label{Wprod}\ee
which is suppressed in our large $\lambda$ regime. Nonzero modes
are more suppressed. We also need to check the
production of open string oscillator modes on
the brane. The masses for these modes are given by the warped
value $m_s(\phi)\sim \phi/\lambda^{1/4}$.  If they were created at
rest, a similar calculation to the one above would yield again
parametric suppression of $\dot m_s/m_s^2\sim 1/\lambda^{1/4}$.
Since the brane is moving with a large velocity
$\sqrt{\lambda}\dot\phi/\phi^2$, the strings created on the brane
have energies boosted to
\be \omega \sim \gamma \sqrt{N_{osc} m_s(\phi)^2+\vec p^2}
\label{fastmodes}\ee
%
%
where $N_{osc}$ is an integer coming from the oscillator level of
the string. These energies are large, and we find that again the
time dependent background does not inject enough energy to create
these states. Therefore our solution is stable against string and
W boson production in the strong coupling regime.

\subsection{Perturbations of $\phi$}

In the remainder of this section, we will analyze the fluctuations
of $\phi$ itself about the solutions we have found. Our main goal
will be to determine in what regime the perturbations are not
dangerous for our solutions above.  These results also pertain to
the spectrum of density perturbations produced in our inflationary
phase.  We will defer a detailed discussion of any predictions of
that model for the CMBR to a later investigation.
We perturb our solutions as, \be
\phi(t)\rightarrow\phi(t)+\alpha(x,t) \nn\ee Expanding in Fourier
modes, $\alpha(x,t)=\alpha_k(t)e^{ikx}$, the equation of motion
for $\alpha$ obtained by expanding the action \eqn{fullact} is,
\be
\ddot{\alpha}_k+\left(\frac{6}{t}+3H\right)\dot{\alpha}_k
+\left(\frac{6}{t^2}+\frac{6H}{t}+\frac{k^2}{\gamma_0^2 a^2t^4}\right)
\alpha_k=0
\nn\ee
where we have used the late time behavior of the
function $\gamma$ defined in \eqn{defgamma} which, for most
of our solutions, is $\gamma\rightarrow t^2/\gamma_0$
for constant $\gamma_0$. In
particular, for the inflationary solution of Case B we have
$\gamma_0\sim mM_p/\sqrt{\lambda}$.

\para
If we analyze this equation of motion in the global case, where
$H=0$, we obtain the results $\delta\phi\propto t^{-2}$ and
$\delta\phi\propto t^{-3}$ for the $k=0$ mode.  For the $k\ne 0$
modes, the perturbations will be further suppressed.  So we see
that the perturbations do not grow relative to the background
solution $\phi\sim\sqrt{\lambda}/t$; i.e. these perturbations are
non-tachyonic.

\para
Let us now analyze the equation with gravity turned on, focusing
on the inflationary solution.  We would like to estimate the
density perturbations that result from the fluctuation
$\delta\phi$ in the field.  In our inflationary phase, we have
\be H = {1\over{\epsilon_0 t}} \label{Hinfl}\ee
where, from \eqn{growing}, we have
$\epsilon_0=(h_1\sqrt{\lambda})^{-1}$. Since we have
$\epsilon_0\ll 1$, in the inflationary phase the terms involving
$H$ and $H/t$ dominate over the terms involving $1/t$ and $1/t^2$
respectively. The relative importance of the $k$-terms depends on
the wavelength under consideration. In general, we therefore have
\be \ddot{\alpha}_k+ \frac{3}{\epsilon_0 t}\dot{\alpha}_k
+\left(\frac{6}{\epsilon_0 t^2}+\frac{c
k^2\lambda}{M_p^2m^2}\frac{1}{a^2t^4} \right)\alpha_k=0
\label{perteom}\ee where $c$ is a constant of order one,
parameterising our ignorance of $\gamma_0$.

\para
For superhorizon fluctuations, defined by $k^2\ll a^2 H^2$, where
$aH$ is the comoving horizon size, we may neglect the spatial
kinetic terms in \eqn{perteom}.
%
%
%
%
%
%
%
This leads to a simple pair of solutions to the classical equation
\eqn{perteom}\ for the modes:
\be
\alpha^{(0)}\sim\frac{\alpha_0(k)}{t^2}
\label{pertsoln0}\ee
and
\be
\alpha^{(1)}\sim\frac{\alpha_1(k)}{t^{3/\epsilon_0}}
\label{pertsoln1}\ee
These modes decay faster at large $t$ than the perturbations of
the scalar field in ordinary slow roll inflation based on
\eqn{usualact}, which go to a constant as the modes cross the
horizon.  There are reasonable regimes of parameters for which all
the subPlanckian $k/a$ modes have the gradient term in
\eqn{perteom}\ suppressed relative to the effective mass squared
term there.

\para
To obtain a prediction for the observed density perturbations, we
must be careful to follow them as they re-enter the horizon at
later times.  For now, let us simply check that the perturbations
do not destabilize our inflationary phase.

\para
During inflation, the density perturbations are simply
\be
\left.{\delta\rho\over\rho}\right|_{k,~during~inflation}\sim
{V^\prime\delta\phi_k\over V}
\label{densinfl}\ee
For our polynomial potentials, this yields $\delta\phi/\phi$. As
we have seen, our perturbations die faster than in ordinary
inflation due to the stronger role for the mass term in
\eqn{perteom}, and they certainly do not compete with $\phi$ in
magnitude. Thus our density perturbations during inflation are
small enough to avoid back reaction. The question of whether one
obtains eternal inflation depends on the magnitude of
$H\delta\phi/{\dot\phi}$, which can be small in our speed of light
phase.
We plan to pursue in future work a detailed study of the
predictions of our inflationary models, as well as an analysis of
reheating and other required features.

\section*{Acknowledgements}
At the risk of sounding like overly-grateful oscar recipients, we
would like to thank B. Acharya, T. Banks, S. Cherkis, N.
Constable, E. Copeland, M. Fabinger, D. Freedman, A. Guth, M.
Headrick, S. Kachru, B. Kors, N. Lambert, A. Linde, H. Liu, J.
Maldacena, E. Martinec, J. Minahan, C. Nunez, H. Ooguri, M. Peskin, J.
Polchinski, M. Schnabl, S. Shenker, W. Taylor, J. Troost and D. Wands for
useful and interesting discussions. And, of course, everyone else
who knows us. We are grateful to the organizers and participants
of the 2003 Benasque string theory workshop for providing a
stimulating environment where this project was initiated. Thanks
also to the organizers and participants of the KITP string
cosmology workshop for hospitality and useful advice on some of
the issues in this paper, and to those of the 2003 International
String School and Workshop in Iran, where these results were first
unveiled. D.T. would like to thank SLAC for hospitality while this
work was completed. E.S. is supported in part by the DOE under contract
DE-AC03-76SF00515 and by the NSF under contract 9870115. This
research was supported in part by the National Science Foundation
under Grant No. PHY99-07949.   D.~T. is supported by a Pappalardo
fellowship, and is grateful to Neil Pappalardo for his generosity.
This work was also supported in part by funds provided by the U.S.
Department of Energy (D.O.E.) under cooperative research agreement
\#DF-FC02-94ER40818.

\small
\parskip=0pt plus 2pt

\end{document}